\documentclass[12pt]{article}

\usepackage{amssymb}
\usepackage{color}
\usepackage{epsfig,amssymb,amsfonts,amsmath,graphicx,dsfont,cite,xfrac}
\usepackage{authblk}
\definecolor{mygray}{gray}{0.5}

\usepackage{cite}
\usepackage[colorlinks=true,linkcolor=blue,citecolor=red]{hyperref}

\newcommand{\be}{\begin{equation}}
\newcommand{\ee}{\end{equation}}
\newcommand{\bea}{\begin{eqnarray}}
\newcommand{\eea}{\end{eqnarray}}

\title{Exactly solvable one-qubit driving fields generated via non-linear equations}

\author[${1,2}$]{Marco Enr\'iquez}
\author[${1}$]{Sara Cruz y Cruz}

\affil[${1}$]{\footnotesize Instituto Polit\'ecnico Nacional, UPIITA, Av. Instituto Polit\'ecnico Nacional 2580 CP 07340 Ciudad de M\'exico, Mexico}

\affil[${2}$]{\footnotesize Tecnologico de Monterrey, Escuela de Ingenier\'ia y Ciencias, Carretera Lago de Guadalupe Km. 3.5, Atizap\'an de Zaragoza, Estado de M\'exico, C.P. 52926, M\'exico}
\date{}
\begin{document}

\maketitle

\begin{abstract}
Using the Hubbard representation for $SU(2)$ we write the time-evolution operator of a two-level system in the disentangled form. This allows us to map the corresponding dynamical law into a set of non-linear coupled equations. In order to find exact solutions, we use an inverse approach and find families of time-dependent Hamiltonians whose off-diagonal elements are connected with the Ermakov equation. The physical meaning of the so-obtained Hamiltonians is discussed in the context of the nuclear magnetic resonance phenomenon.
\end{abstract}


\section{Introduction}

The dynamical manipulation of two-level systems has been a long-standing issue in quantum mechanics \cite{Rab37,Zen32,Ros32}. In recent times the problem has acquired more relevance due to its potential applications in quantum computing and quantum information \cite{Nie00,Ger09,Fuc09}. Eventhough it is widely acknowledged that the number of exactly solvable cases is very limited, some analytical solutions have been recently reported in the literature \cite{Fuc09,Bez11,Gan10,Xie10,Vit07,Bar12,Mes14}. 
The aim of the present paper is to propose a method to obtain new families of analitically solvable driving fields by requiring that the corresponding time-evolution operator exactly factorizes as a product of exponentials whose arguments are proportional to the generators of the $su(2)$ algebra.
This procedure is related to the well-known Wei-Norman theorem \cite{Wei63}
which has been used to study the dynamics of systems with $SU(2)$ and $SU(1,1)$ symmetries \cite{Dat88n,Dat86j,Dat87,Dat88}, e.g., the interaction of two-level atoms with electromagnetic radiation \cite{Pra11}, field modulation in nuclear magnetic resonance \cite{Cam89}, propagation and perfect transmission in three-waveguide axially varying structures \cite{Rod14}, the time-evolution of harmonic oscillators with time-dependent both mass and frequency \cite{Dat86p,Lo90} among others.
%
%
We are interested in the dynamics of systems whose Hamiltonians can be written in the form
\begin{equation}\label{h21}
 H(t)=\frac\Delta 2J_0+V(t) J_++{\overline V}(t) J_-,
\end{equation}
where the operators $\{J_0,J_\pm\}$ are the generators of the $su(2)$ algebra and $\Delta \in \mathbb{R}$, so that the operator (\ref{h21}) is Hermitian. For the 2D representation one obtains the Hamiltonian of a driven two-level system (or {\it qubit}). The complex function $V$ describes the interaction of the system with the external field and it will be referred to as the driving field. 
The Wei-Norman theorem allows us to write the corresponding time-evolution operator in the so-called disentangled form
\begin{equation}\label{utsu2}
 U(t)=\displaystyle e^{\alpha(t) J_+}e^{\Delta f(t) J_0}e^{\beta(t) J_-},
\end{equation}
where $\alpha, f$ and $\beta$ are the complex-valued factorizing functions to be determined. In Ref. \cite{Enr17} we show that these can be found in terms of the solutions of the parametric oscillator-like equation $\varphi''(t) + \Omega^2 (t) \varphi(t) =0$, with the time-dependent {\it frequency} given by
\begin{equation}\label{tdfreqv}
 \Omega^2(t) = -\frac14 \left[\frac d{dt} \ln {\overline V}(t)-i \Delta \right]^2 + \frac12 \frac{d^2}{dt^2} \ln {\overline V} (t)+ \vert V(t)\vert ^2.
\end{equation}
Hence, one can realize that the core difficulty to express $U$ in the form (\ref{utsu2}) lies in the fact that the solutions to the parametric oscillator-like equation can be explicitly written only in a restricted number of cases. 
Therefore, we address the disentangling problem using an inverse approach, 
where some aspects on the dynamics are prescribed and then the interactions are found. For instance, in Ref. \cite{Fer97} some families of exact solvable driving fields for spin-1/2 systems are obtained by requiring the spin to accomplish particular trajectories in the projective space. This approach has been also widely used 
in finding analytical solutions of a two-level system driven by a single-axis control field
\cite{Gan10,Bar12,Mes14}.
In our case, we require the time-evolution operator to have the exact form (\ref{utsu2}) and look for different types of the function $V$ that allow us to exctly construct the factorizing functions $\alpha$, $f$ and $\beta$. In other words, we should solve equation (\ref{tdfreqv}) with a function $\Omega$ for which the solutions to the parametric oscillator-like equation are known. This demands to solve the highly nonlinear equation (\ref{tdfreqv}), however it will be shown that with the proper change of variable this can be transformed into the Ermakov equation related to a parametric oscillator-like equation. Thus, our procedure constitute a mechanism to generate families of exactly solvable single qubit driving fields for which the disentangling problem is exactly solvable.
%
%

The paper is organized as follows. In Section~\ref{dirapp} the preliminaries and the direct problem are reviewed. The main result is reported in Sec.~\ref{invapp}. In addition, the physical meaning of our results is discussed in Sec.~\ref{physmod}. Some concrete examples are analyzed in Sec.~\ref{examples} where we show that our results generalize the case of a circularly polarized field and discuss some aspects of the dynamics such as the time-evolution of population inversion. Finally, the conclusions and perspectives of our work are presented in Sec.~\ref{conc}.

\section{The direct approach}\label{dirapp}

We first present some properties of the Hubbard operators (also called X-operators) to construct the corresponding representation of the $su(2)$ algebra. The X-operators \cite{Enr13} form 
a set of $n^2$ two-labeled operators $\{X^{i,j}\}_{i,j=1}^n$ 
fullfiling the properties
\begin{enumerate}
\item $X^{i,j} X^{k,m} = \delta_{j,k} X^{i,m}$ (multiplication rule)
 
 \item\label{comprop}
 $\displaystyle \sum_{k} X^{k,k} = \mathbb I$ (completeness)
 
 \item
 $(X^{i,j})^{\dagger} = X^{j,i}$ (non-hermiticity)
\end{enumerate}
%
The most elementary case arises with $n=2$ and can be used to describe the time-evolution of a two-level system. The algebraic properties of the Hubbard operators allow to deal with some calculations in a simpler way, as well as to extend our formalism to the cases of qudits and multipartite systems \cite{Enr17}. For the two-level system with eigenstates $\vert p\rangle$ and $\vert q \rangle$ corresponding to the eigenvalues $\epsilon_p$, $\epsilon_q$, respectively the four operators defined as
\begin{equation}\label{klop}
 X_2^{k,\ell}:=\vert k \rangle\langle \ell\vert, \quad k,\ell=p,q.
\end{equation}
constitute a representation of the set of Hubbard operators. Besides, a simple calculation shows that the action of (\ref{klop}) on the basis vectors reads
\begin{equation}
 X_2^{k,\ell} \vert n\rangle = \delta_{\ell,n} \vert k\rangle, \quad k,\ell, n=p,q.
\end{equation}
On the other hand, it is easy to verify that the three operators
\begin{equation}
 J_0:=\frac12 (X^{p,p}_2-X^{q,q}_2), \quad J_+=(J_-)^\dagger:=X^{p,q}_2,
\end{equation}
constitute a representation of the $su(2)$ algebra. Remark that, eventhough we are using the lowest representation with $j=1/2$, our results are equivalent for any value of $j$ once the generators are written in the proper realization (See  \cite{Enr13} for details).
The Hamiltonian (\ref{h21}) in terms of the X-operators reads
\begin{equation}\label{h2}
 H_2(t) = \frac\Delta2(X^{p,p}_2-X^{q,q}_2) + V(t) X^{p,q}_2+{\overline V}(t)  X^{q,p}_2. 
\end{equation}
A discussion on the physical interpretation of this operator is given in Sec.~\ref{physmod}.
 
We will address the problem of finding the time-evolution operator for the Hamiltonian (\ref{h2}) by solving the dynamical law
\begin{equation}\label{timeeq}
i \frac{d U(t)}{dt}= H_2(t) \cdot U(t)
\end{equation}
with the initial condition $U(0)=\mathbb{I}$. Using the Wei-Norman theorem, the operator (\ref{timeeq}) can be written as
%
\begin{equation}\label{ut}
 U(t) = \exp[\alpha(t) X^{p,q}_2] \exp [\Delta f(t) J_0] \exp[\beta(t) X^{q,p}_2],
\end{equation}
where the complex factorization functions $\alpha, f$ and $\beta$ are to be determined provided that $\alpha(0)=f(0)=\beta(0)=0$. These last are consequence of the initial condition for $U$. The action of the time-evolution operator (\ref{ut}) on the basis vectors can be straightforwardly computed using the following relations
%
\begin{equation}
\begin{array}{l}
\exp[\alpha(t) X^{p,q}_2] = 1+\alpha(t) X^{p,q}_2, \quad \exp[\beta(t) X^{q,p}_2] = 1+\beta(t) X^{q,p}_2,\\[1em]
\exp\left[\frac{\Delta f(t)}2 X^{p,p}_2\right] = X_2^{q,q}+e^{\Delta f(t)/2} X^{p,p}_2 , ~\exp\left[\frac{-\Delta f(t)}2 X^{q,q}_2\right] = X_2^{p,p}+e^{-\Delta f(t)/2}  X^{q,q}_2,\\[1em]
\exp\left[\Delta f(t) J_0\right] = e^{\Delta f(t)/2} X_2^{p,p} + e^{-\Delta f(t)/2} X_2^{q,q}.
\end{array}
\end{equation}
Thus, if the qubit is initially, for instance, in the upper-level energy state $\vert p\rangle$, the state of the system at any time reads
\begin{equation}\label{psipt}
 \vert \psi(t)\rangle = e^{-\Delta f(t)/2} [e^{\Delta f(t)}+\alpha(t)\beta(t)]\vert p \rangle + e^{-\Delta f(t)/2} \beta(t) \vert q\rangle.
\end{equation}
On the other hand, in Ref. \cite{Enr17} we show that the solutions to the evolution problem (\ref{timeeq})-(\ref{ut}) is given by
\begin{equation}\label{falpha}
\alpha(t) = \displaystyle\frac{e^{-i \Delta t}}{R(t)}\left[ \frac{\varphi'(t)}{\varphi(t)} +\frac 12 \frac {R'(t)}{R(t)}\right],
\end{equation}
\begin{equation}\label{ff}
 \Delta f(t) = -2 \ln \left[\frac{\varphi(t)}{\varphi_0} \right] - \ln \left[\frac{R(t)}{R_0}\right]-i\Delta t, \quad \varphi_0=\varphi(0), \quad R_0=R(0),
\end{equation}
and
\begin{equation}\label{fbeta}
 \beta(t) = R_0 \int_0^t \frac{ds}{\varphi^2(s)},
\end{equation}
where the function $R(t) = -i e^{-i \Delta t} ~{\overline V}(t)$, which will be also refered to as the driving field, has been introduced to shorten the notation and  the function $\varphi$ is a solution of the parametric oscillator-like equation
\begin{equation}\label{parosc}
 \varphi''(t) + \Omega^2(t) \varphi(t) =0,
\end{equation}
with time-dependent frequency given by
\begin{equation}\label{tdfreq}
 \Omega^2(t) = -\frac14 \left[\frac d{dt} \ln R(t)\right]^2 + \frac12 \frac{d^2}{dt^2} \ln R(t)+ \vert R(t)\vert ^2.
\end{equation}
This equation is of course equivalent to (\ref{tdfreqv}). Besides, the initial condition $\alpha(0)=0$ state that
\begin{equation}\label{inicon}
\lim_{t\rightarrow 0} \frac1{R(t)} \left[\frac{\varphi'(t)}{\varphi(t)}+\frac 12 \frac {R'(t)}{R(t)}\right]=0. 
\end{equation}
Without loss of generality may set $\varphi_0=\varphi(0)=1$, as the predictions will not depend on this initial value. We emphize that given $V(t)$ (or equivalently $R(t)$) the problem of finding the time-evolution operator $U(t)$ for the Hamiltonian (\ref{h2}) is turned into solving equation (\ref{parosc}). 
As it was mentioned before, the number of exactly solvable cases is limited since the form of the frequency (\ref{tdfreq}) is in general non-trivial. 
Some exactly solvable cases are $V(t)={\rm const}$ \cite{Dat86j}, the circularly polarized field: $V(t)\propto e^{-i\delta t}$ \cite{Enr17} and the hyperbolic secant pulse: $V(t)\propto {\rm sech}(\sigma t)$ \cite{Cam89}, to mention some.
In this work, to obtain analytical solutions, we will address the problem using an inverse approach: we look for the driving field $V(t)$ that gives rise to a given time-dependent frequency (\ref{tdfreq}) for which the solutions of (\ref{parosc}) are known.

\section{The inverse approach}\label{invapp}
In this section we obtain several families of exactly solvable one-qubit time-dependent Hamiltonians by constructing driving fields $R(t)$, as solutions of (\ref{tdfreq}), for a given function $\Omega(t)$. First note that this equation is not of the Riccati-type in the function $\ln R(t)$ because of the term $\vert R(t)\vert^2$. Therefore, in order to transform this equation into a known one, let us consider the function $\gamma$ defined by
%
%
\begin{equation}
 \gamma(t)=\frac{d}{dt}\ln R(t).
\end{equation}
It follows immediately that
\begin{equation}\label{ansatzR}
 R(t) = R_0\exp\left[\int_0^t \gamma(s)~ds\right].
\end{equation}
In general $\gamma$ a complex-valued function that can be splitted into its real and imaginary parts as $\gamma(s)=\gamma_1(s)+i\gamma_2(s)$. 
The corresponding initial conditions for $\gamma$ are determined through the initial conditions on $R(t)$
\begin{equation}\label{icgamma}
 \gamma_1(0)={\rm Re}\left[\frac{R'_0}{R_0}\right], \quad \gamma_2(0)={\rm Im}\left[\frac{R'_0}{R_0}\right],
\end{equation}
where $R_0'=R'(0)$. We will restrict ourselves to the case in which the function $\Omega$ is real-valued. Under this assumption, the substitution of (\ref{ansatzR}) into (\ref{tdfreq}) leads to the set of equations
\begin{equation}\label{refreq}
-\frac14\gamma_1^2+\frac14\gamma_2^2+\frac12\gamma_1'+\vert R_0\vert^2 \exp\left(2\int_0^t\gamma_1 ds\right) = \Omega^2,
\end{equation}
%
%
and
\begin{equation}\label{g1g2}
 \gamma_1=\frac{\gamma_2'}{\gamma_2}=\frac d{dt}\ln \gamma_2.
\end{equation}
%
%
A pair of equations similar to (\ref{refreq}) and (\ref{g1g2}) was analyzed by Rosas-Ortiz {\it et al.} in the context of the Schr\"odinger equation with complex-valued potentials \cite{Ros15}. The main difference with our case is the last term in the left-hand side of equation (\ref{refreq}), which makes the equation highly non-linear. Nevertheless, it turns out that this equation can be reduced to an Ermakov equation by the introduction of a new real-valued function $\mu$ such that
%
\begin{equation}\label{g1mu}
 \gamma_1 = -\frac d{dt}\ln \mu^2.
\end{equation}
%
Inserting this expression in (\ref{g1g2}) we obtain
\begin{equation}\label{g2mu}
 \gamma_2=\frac\lambda{\mu^2},
\end{equation}
where $\lambda$ is an integration constant 
and $\mu$ satisfies
\begin{equation}\label{ermakovmu}
 \mu''(t)+ \Omega^2(t)\mu(t) =\frac{\Omega_0^2}{\mu^3(t)}, \quad \Omega_0^2=\lambda^2\left[\frac{\vert R_0\vert^2}{\gamma_2^2(0)}+\frac14\right].
\end{equation}
%
According to equations (\ref{icgamma}) the three parameters $\mu_0$, $\mu_0'$ and $\lambda$ are related through
%
\begin{equation}
 \frac{\mu_0'}{\mu_0} =-\frac12 {\rm Re} \left[\frac{R_0'}{R_0}\right], \quad \frac{\lambda}{\mu_0}={\rm Im} \left[\frac{R_0'}{R_0}\right],
\end{equation}
however, without loss of generality we may fix $\mu_0=1$.
Moreover, the constant
%
\begin{equation}
  \Omega_0=\left[\vert R_0\vert^2+\frac{\lambda^2}4\right]^{1/2}, \quad {\rm where } \quad \lambda = \gamma_2(0).
\end{equation}
will be referred as to the generalized Rabi-like frequency. Finally, inserting the functions $\gamma_1$ and $\gamma_2$ in (\ref{ansatzR}) we obtain the driving fields of the solution of the Ermakov equation (\ref{ermakovmu})
\begin{equation}
 R(t) = \frac{R_0}{\mu^2(t)}\exp\left[i\lambda \int_0^t \frac{ds}{\mu^2(s)}\right].
\end{equation}
This function defines a family of exactly solvable time-dependent Hamiltonians  of the form (\ref{h2})
with $V(t)=-ie^{-i\Delta t}{\overline R}(t)$. 

It is worthwhile to stress that the solutions to equation (\ref{ermakovmu}) can be constructed provided that the soltutions of (\ref{parosc}) are known. Indeed, it can be shown that if $\varphi_1$ is a particular solution of such equation the function $\mu$ is given by \cite{Ros15,Erm80}:
\begin{equation}\label{esol}
 \mu^2(t) = \frac{\Omega_0^2 \varphi_1^2(t)}{c_1}+c_1 \varphi_1^2(t)\left[c_2+\int\frac{dt}{\varphi_1^2(t)}\right]^2,
\end{equation}
where $c_1$ and $c_2$ are integration constants \cite{Erm80, Pin50}.
Moreover, for the so-obtained Hamiltonians the time-evolution operator $U(t)$ is readily written from equation (\ref{ut}) with the factorization functions $\alpha, f$ and $\beta$
\begin{equation}
 \alpha(t)=\frac{\mu^2(t)}{R_0} \exp\left[-i\Delta t-i\lambda \int_0^t \frac{ds}{\mu^2(s)}\right] \left[\frac{\varphi'(t)}{\varphi(t)}-\frac{\mu'(t)}{\mu(t)}+\frac{i\lambda}{2\mu^2(t)}\right],
\end{equation}
\begin{equation}
\beta(t)=R_0 \int_0^t \frac{ds}{\varphi^2(s)},
\end{equation}
\begin{equation}
 \Delta f(t) = \ln\left[\frac{\mu^2(t)}{\varphi^2(t)}\right] - i\lambda \int_0^t \frac{ds}{\mu^2(s)}-i\Delta t.
\end{equation}
%

%

\subsection{Periodic interactions}
Of particular interest are the driving fields fulfilling the condition
$R(t)=R(t+\tau_p)$. Remark, that a periodic function $\mu$ does not assure the periodicity of $R$. Indeed, if the period of the function $\mu$ is $\tau$ it can be shown that
%
%
%
\begin{equation}\label{rper}
 R(t+\tau)=\exp\left[i\lambda \int_0^\tau \frac{ds}{\mu^2(s)}\right] R(t).
\end{equation}
%
From this last expression it is clear that for $R$ to be a periodic function we should adjutst $\lambda$ and the parameters involved in the function $\mu$ 
in a very restrictive form. However, note that for a natural number $p$ the following expression holds
\begin{equation}\label{ptau}
R(p \tau)=\exp\left[ip\lambda \int_0^\tau \frac{ds}{\mu^2(s)}\right] R_0.
\end{equation}
This parameter $p$ gives versatility to manipulate the period of the function $R$. Indeed, by requiring that
\begin{equation}\label{pcond}
p \lambda \int_0^\tau\frac{ds}{\mu^2(s)}\equiv 0 \quad ({\rm mod}~ 2\pi),
\end{equation}
it is assured that $R$ is a periodic function with period
$\tau_p=p\tau$.
\section{A physical model}\label{physmod}
\begin{figure}[t]
\begin{center}
\includegraphics[scale=0.45]{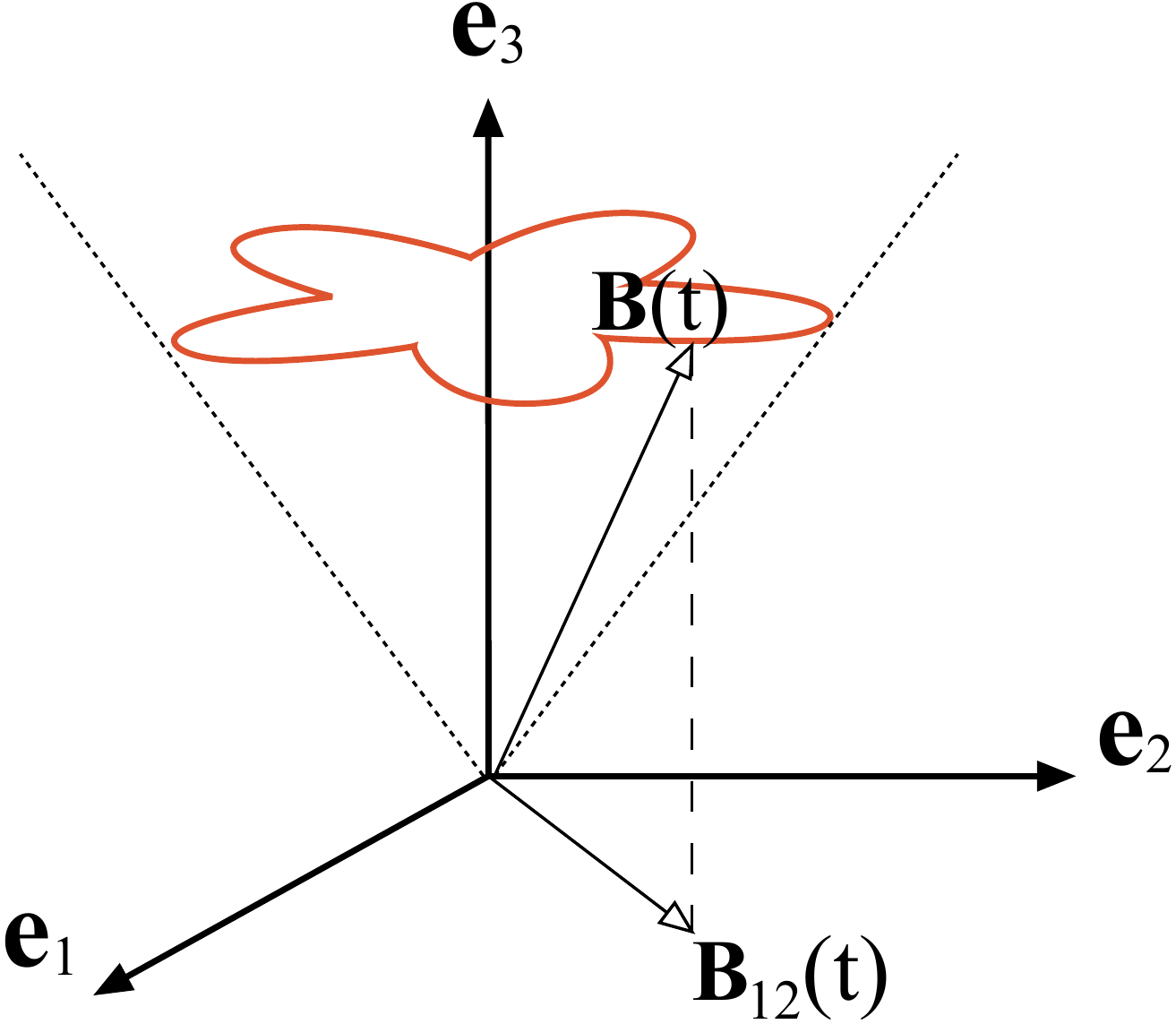}
\caption{\footnotesize Schematical representation of the magnetic field ${\bf B}$. The transversal component ${\bf B}_{12}$ rotates around the axis ${\bf e}_3$ in the plane $z=B_3$. The amplitude of the driving field is time-dependent and describes a trajectory that in general is not closed. The circularly polarized field is a particular case of this expression when the amplitude is constant.}
\label{field}
\end{center}
\end{figure}

In this section we consider a quantum system whose Hamiltonian can be written in terms of the Hubbard operators given in (\ref{h2}).  Consider a spin-1/2 particle with magnetic moment ${\bf m} =\mu_B g_\ell {\vec \sigma}$ in an external magnetic field ${\bf B}(t)$, where $\mu_B$ and $g_\ell$ stand for the Bohr magneton and the Land\'e factor, respectively, and ${\vec \sigma}=(\sigma_1,\sigma_2,\sigma_3)$. The corresponding Hamiltonian reads
\begin{equation}\label{hmb}
H=-{\bf m}\cdot {\bf B} = b {\vec \sigma}\cdot {\bf B}, 
\end{equation}
where $b=-g_\ell \mu_B$ is a constant 
We pay attention to the case where the field precesses around a fixed axis,  say the axis ${\bf e}_3$. Thus, the magnetic field is written as ${\bf B}={\bf B}_{12}(t)+B_3{\bf e}_3$, where $B_3$ is a constant and the driving field ${\bf B}_{12}(t) = B_1(t){\bf e}_1+B_2(t){\bf e}_2$ is used to manipulate the state of the spin.
This field is represented schematically in Fig~\ref{field}. In Hubbard notation the Hamiltonian (\ref{hmb}) is expressed as
\begin{equation}\label{h2h}
 H=\frac{b B_3}{2} ({X^{p,p}_2-X^{q,q}_2})+\frac b2[B_1(t)-iB_2(t)]{X^{p,q}_2}+\frac b2 [B_1(t)+iB_2(t)] {X^{q,p}_2}.
\end{equation}
%
Thus, with the identification $V(t)= b[B_1(t)-i B_2(t)]/2$ and $\Delta = b B_3$, the Hamiltonian (\ref{h2h}) takes the form (\ref{h21}). Note that the real and imaginary parts of the function $V$ are proportional to the component 1 and 2 of the field, respectively. This justifies the name given to the function $V$. In particular, if the transverse field has a constant amplitude $B_0$, we get the well-known case of a circularly polarized field, as the transverse vector ${\bf B}_{12}$ describes a circle in a plane transverse to ${\bf e}_3$. We may take $B_1(t)=B_0\cos \omega t$ and $B_2(t)=B_0\sin \omega t $, where $\omega$ is the oscillation frequency. Hence $V(t)=\frac b2 B_0 e^{-i\omega t}$ and the solutions to the factorization problem (\ref{ut}) are easily constructed. We obtain $R(t)=-i g e^{i \delta t }$ where $g=\frac b2 B_0$ is a real constant, and the detuning $\delta = \omega -\Delta$. The factorization functions are (See \cite{Enr17}):
\begin{equation}\label{afbcir}
\begin{array}{ccc}
 \alpha(t) = \displaystyle \frac{-2 i g e^{-i\omega t} \sin(\Omega_0 t)}{2 \Omega_0 \cos (\Omega_0 t)-i\delta \sin (\Omega_0 t)},\\[1em]
 \Delta f(t) = \displaystyle -2\ln \left[ \cos (\Omega_0 t)-\frac{i\delta}{2 \Omega_0}\sin(\Omega_0 t)\right]-i\omega t,\\[1em]
 \beta(t) = \displaystyle \frac{-2i {g} \sin(\Omega_0 t)}{2 \Omega_0 \cos (\Omega_0 t)-i\delta \sin (\Omega_0 t)},
\end{array}
\end{equation}
where $\Omega_0^2=g^2+\delta^2/4$ is the Rabi frequency. Besides, if the system is initially in the upper-level energy state, \emph{i.e.}, $\vert \psi(0) \rangle= \vert p\rangle$, the state of the system at any time can be computed as $\vert \psi(t)\rangle = U(t) \vert p \rangle$ to yield
\begin{equation}
 \vert\psi(t)\rangle = e^{-i\omega t/2}\left[\cos (\Omega_0 t) +\frac{i\delta}{2\Omega_0}\sin(\Omega_0 t)\right]\vert p\rangle -\frac{i\,g}{\Omega_0} e^{i\omega t/2} \sin (\Omega_0 t)\vert q\rangle.
\end{equation}
In addition, the atomic population inversion $P=\vert c_p\vert^2-\vert c_q\vert^2$ is also computed:
\begin{equation}\label{popcp}
P(t) = \frac{ g^2}{\Omega_0^2}\cos (2 \Omega_0 t)+\frac{\delta^2}{4 \Omega_0^2}.
\end{equation} 
%
%

%

\section{Examples}\label{examples}

In this section we explicitly obtain several families of one-qubit time-dependent Hamilitonians for a given function $\Omega$. This function should be chosen in such a way the solutions to the parametric oscillator equation (\ref{parosc}) are known. 
%
The construction of the driving field $R(t)$ through equation (\ref{tdfreq}) involve two arbitrary parameters, namely $R_0$ and $R_0'$. In this section we will discuss the case for which $[\ln R(0)]'=i\delta$, with $\delta\in \mathbb{R}$, leading to $[\ln \mu(0)]'=0$. We may also write $R_0=-i{\overline g}$, where $g$ may be in general a complex constant. Observe that these conditions are analogue to those of the problem of a qubit interacting with a circularly polarized field discussed in the previous section. 
%
%

We will consider that the frequency $\Omega$ is a non-negative constant $\Omega_1$. In this case two important limits are of interest. In the first one $\Omega_1=0$, which leads to a decaying driving field. The second one arises as $\Omega_1\rightarrow \Omega_0$ and 
corresponds to the problem of one qubit interacting with a single mode circularly polarized field. 
\subsection{Case $\Omega(t)=0$}

\begin{figure}[t]
\begin{center}
\includegraphics[scale=0.41]{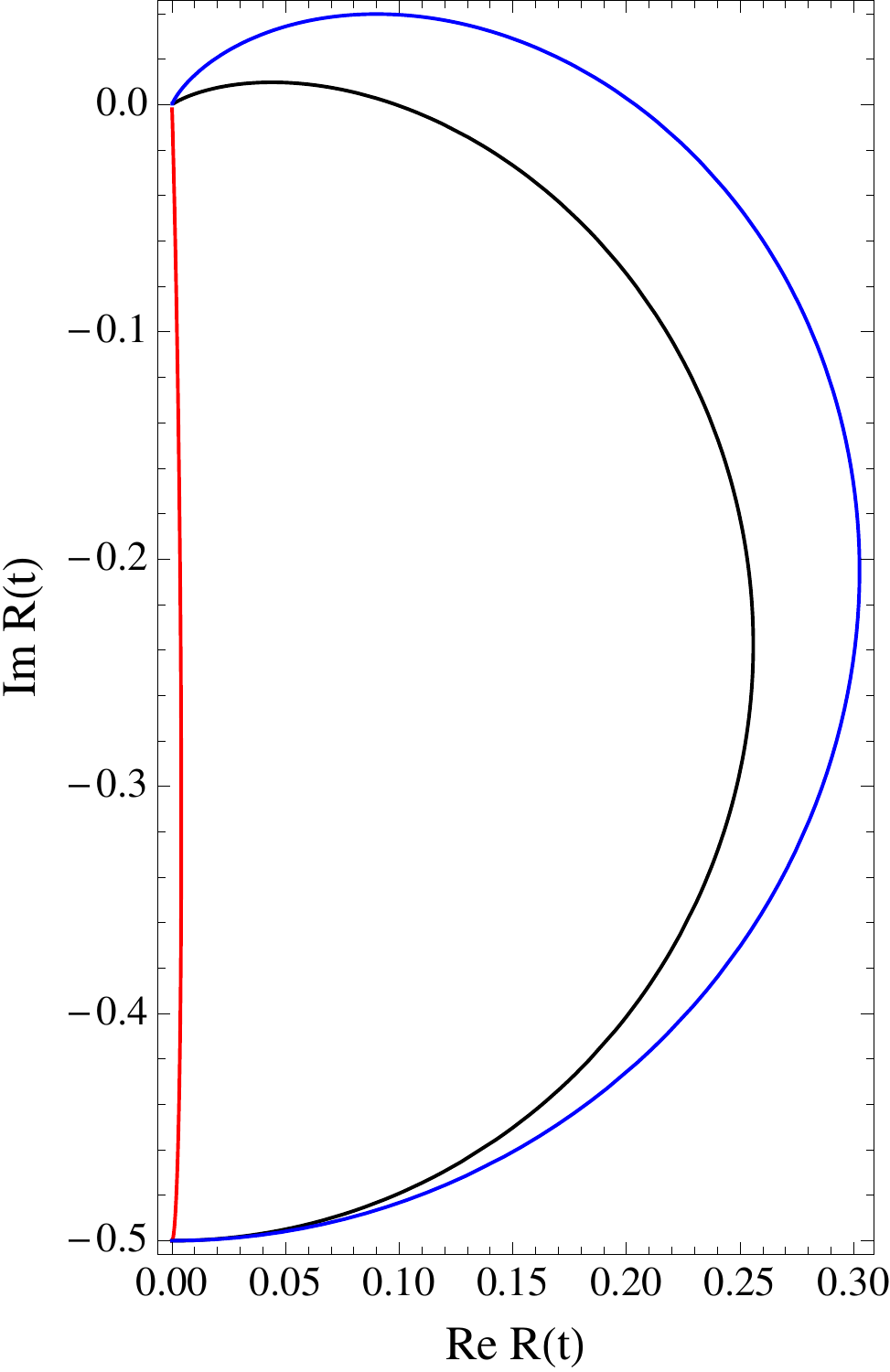}\quad\quad\quad\qquad
\includegraphics[scale=0.5]{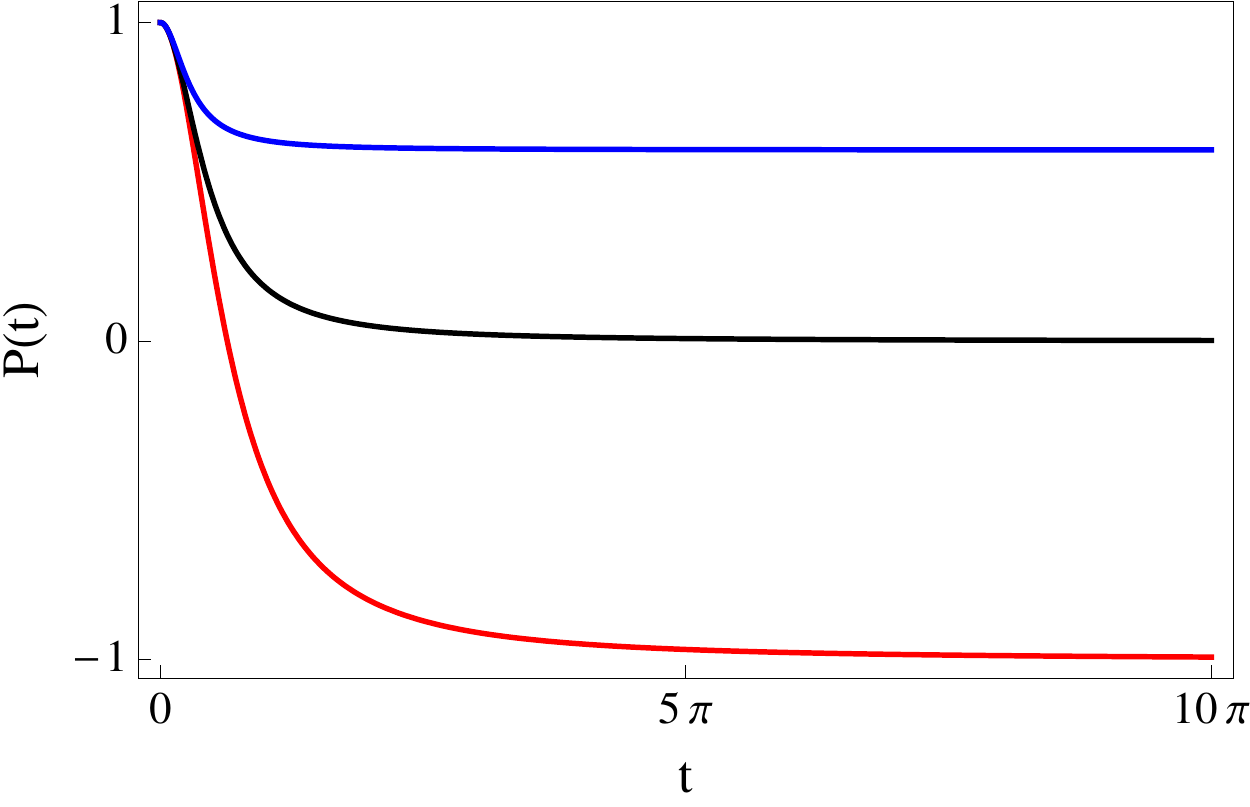}
\caption{\footnotesize (Right) The interaction term (\ref{Ro0}) in the complex plane. (Left) Population inversion as function of time. The parameters in both cases are: $g=0.5$, $\delta=0.01$ (red), $\delta=1$ (black) and $\delta =2$ (blue)}
\label{PRo0}
\end{center}
\end{figure}

On one hand, note that when the frequency $\Omega(t)$ vanishes the parametric oscillator equation (\ref{parosc}) becomes
%
$\varphi''(t) =0,$
%
whose solution meeting the initial conditions (\ref{inicon}) is $ \varphi(t)=-i\delta t/2+1$. Next, the function $\mu$ can be found by taking $\varphi_1(t)=t$ in (\ref{esol}). 
The initial conditions $\mu(0)=1$ and $[\ln \mu(0)]'=0$ lead to $\mu^2(t)=\Omega_0^2 ~t^2+1$ and to the corresponding driving field 
\begin{equation}\label{Ro0}
 R(t)=\frac{-i {\overline g}}{\Omega_0^2 t^2+1} \exp\left[i\frac\delta{\Omega_0} \arctan(\Omega_0 t)\right], \quad \Omega_0^2 = \vert g\vert^2+\frac{\delta^2}4.
\end{equation}
The factorizing functions $\alpha, \beta$ and $f$ can be explicitly constructed to yield
\begin{equation}
 \alpha(t) = \frac{-2igt}{-i\delta t+2} \exp\left[-i\Delta t-i\frac\delta{\Omega_0} \arctan(\Omega_0 t)\right], \quad \beta(t) = \frac{-2i {\overline g}t}{-i\delta t+2},
\end{equation}
and
\begin{equation}
 \Delta f(t)=\ln\left[\frac{4\Omega_0^2t^2+4}{(-i\delta t+2)^2}\right]+i\frac\delta{\Omega_0}\arctan(\Omega_0 t)-i\Delta t.
\end{equation}
According to equation (\ref{psipt}) the time-evolution of the state $\vert \psi \rangle$ can be computed in terms of the factorization functions as
\begin{equation}
\begin{array}{ll}
\displaystyle \vert\psi(t)\rangle = \frac{(i \delta t+2)e^{-i\Delta t/2}}{2 \mu(t)}  \exp\left[-i\frac\delta{2\Omega_0} \arctan(\Omega_0 t)\right]\vert p\rangle\\[1em]
\hspace*{7cm} \displaystyle-i\frac{{\overline g}te^{i\Delta t/2} }{\mu(t)} \exp\left[i\frac\delta{2\Omega_0} \arctan(\Omega_0 t)\right]\vert q\rangle.
 \end{array}
\end{equation}
Hence the population inversion $P$ is calculated as function of time
\begin{equation}
 P(t)=\frac{(\delta^2/4-\vert g\vert^2) t^2+1}{\Omega_0^2t^2 +1}.
\end{equation}
Fig.~\ref{PRo0} shows the function $R$ in the complex plane for several values of the parameters $g$ and $\delta$. Remark that the real and imaginary parts can be considered as the transversal components of a control magnetic field.
It can be seen from (\ref{Ro0}) that as $t\rightarrow\infty$, we have $R(t)\rightarrow0$. This fact is noted in the population inversion $P$, which tends to fixed a value as time grows (see Fig.~\ref{PRo0}). Indeed
\begin{equation}
 P(t\rightarrow\infty)=\frac{\delta^2/4-\vert g\vert^2}{\Omega_0^2}.
\end{equation}

\subsection{Case $\Omega(t)=\Omega_1$}

A more general instance arises if the frequency is a real constant $\Omega_1\neq0$. With this choice the parametric oscillator-like equation reduces to the conventional Newtonian harmonic oscillator equation
\begin{equation}
 \varphi''(t) + \Omega_1^2 ~\varphi(t) =0,
\end{equation}
whose solution is a linear combination of the harmonic functions $\cos(\Omega_1t)$ and $\sin(\Omega_1 t)$. The function that meets the initial conditions (\ref{inicon}) reads
\begin{equation}\label{phio}
 \varphi(t)=\cos(\Omega_1 t)-\frac{i\delta}{2\Omega_1}\sin(\Omega_1t).
\end{equation}
On the other hand, the solution to the corresponding Ermakov equation is given by the function (\ref{esol}) with $\varphi_1(t)=\cos (\Omega_1t)$. It is obtained
\begin{equation}\label{muo}
 \mu^2(t)=\cos^2(\Omega_1t)+\kappa^2 \sin^2(\Omega_1 t), \quad \kappa=\Omega_0/\Omega_1.
\end{equation}
The corresponding driving field is given by 
\begin{equation}\label{rto}
R(t)=\frac{-i \, {\overline g}\,e ^{i\delta\, \eta(t)}}{\cos^2(\Omega_1t)+\kappa^2 \sin^2(\Omega t)}
\end{equation}
with
\begin{equation}\label{ieta}
 \eta(t)=\int_0^t \frac{ds}{\cos^2(\Omega_1s)+\kappa^2 \sin^2(\Omega s)}\,.
\end{equation}
The interaction term (\ref{rto}) has the form of a precessing field with an oscillating amplitude.
Note that a circularly polarized field is a particular case of our driving fields. Indeed, taking $\kappa=1$ we have $\mu^2(t)=1$ and hence $\eta(t)=t$. Therefore $R(t)=-i{\overline g} e^{i \delta t}$, which is the field discussed in the section \ref{physmod}.

\begin{figure}[t]
\begin{center}
\begin{tabular}{ccc}
\includegraphics[scale=0.45]{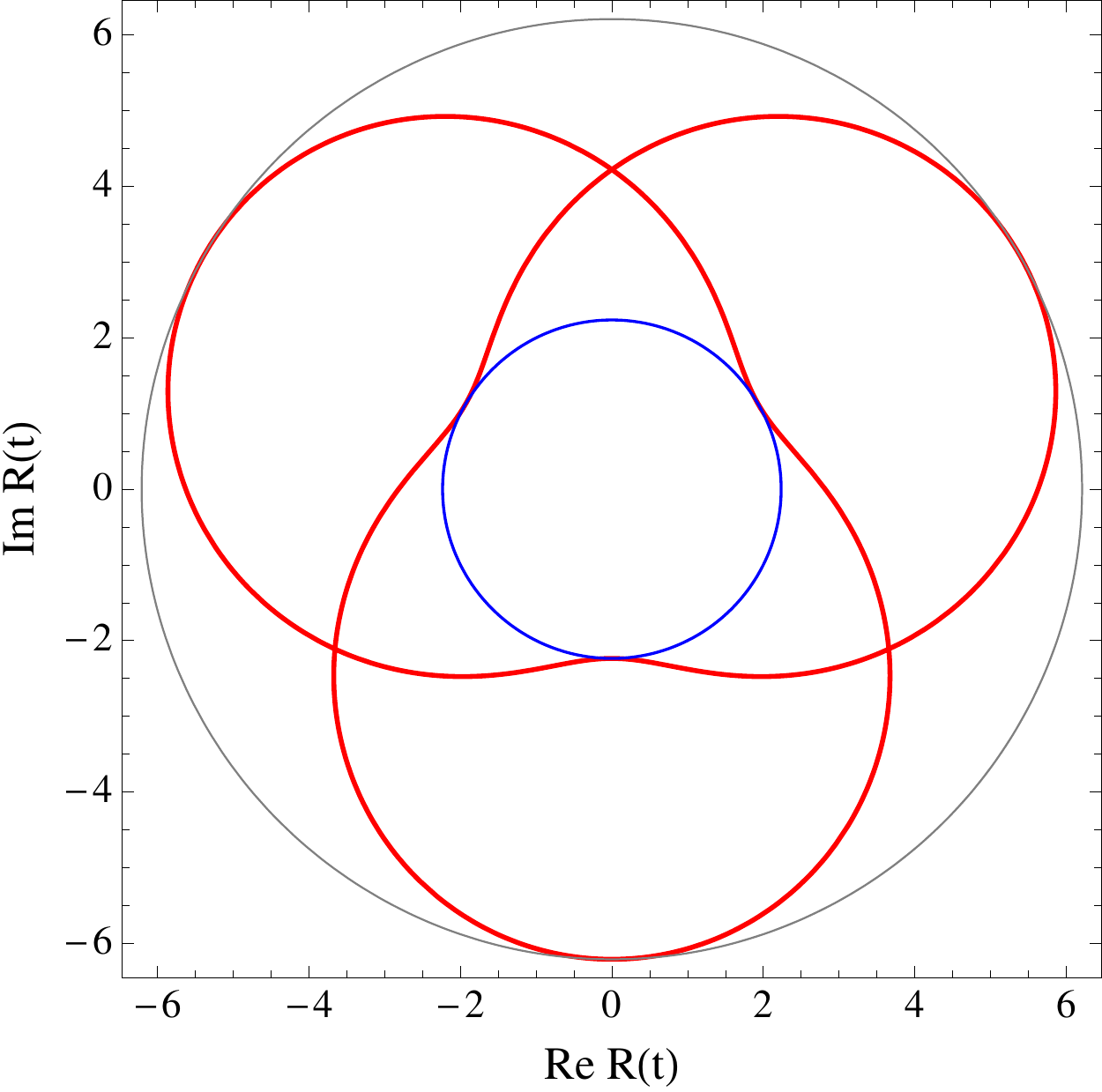}&\hspace*{0.4cm}&\includegraphics[scale=0.45]{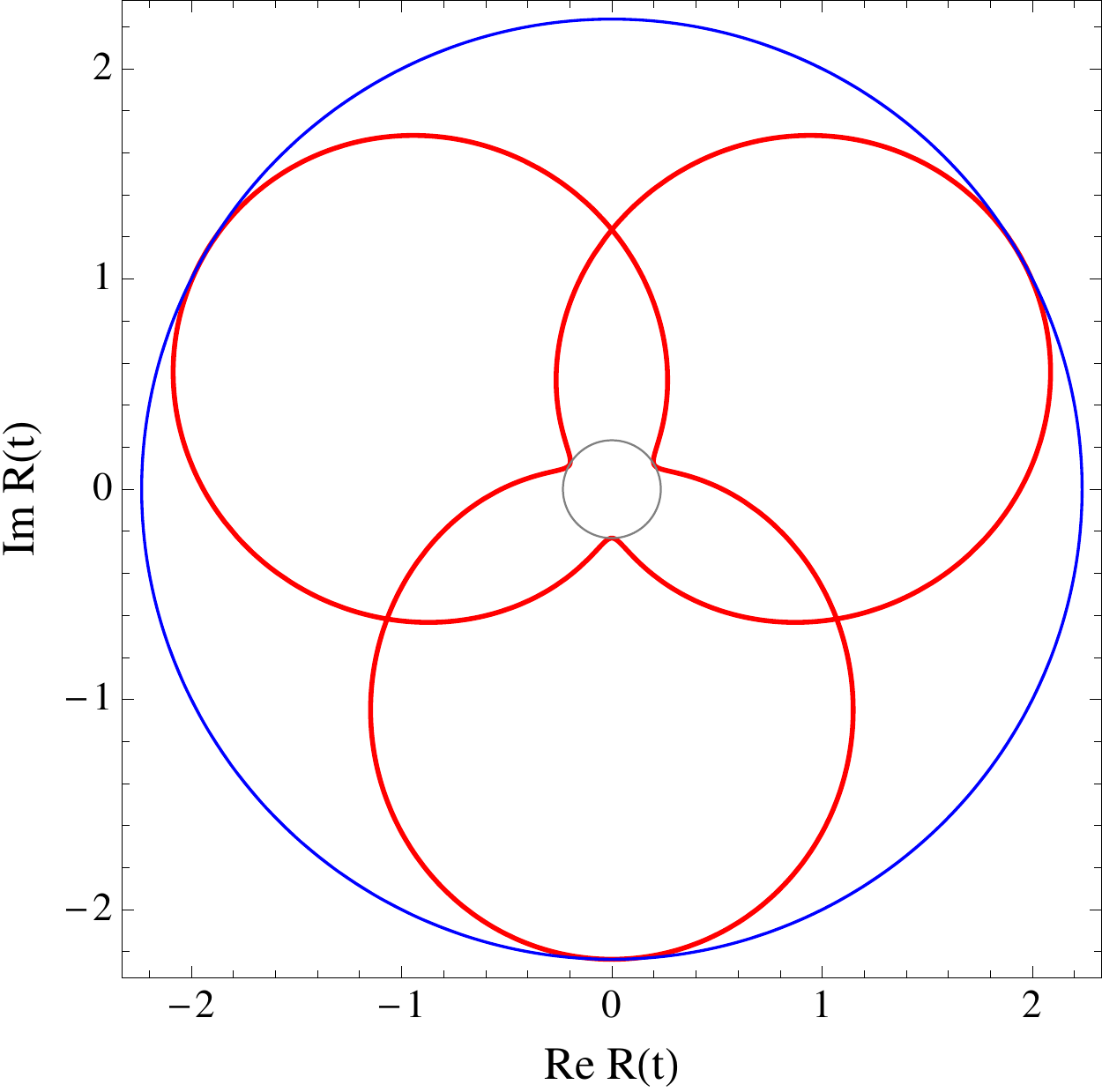}\\
a) & & b)\\[1em]
\includegraphics[scale=0.45]{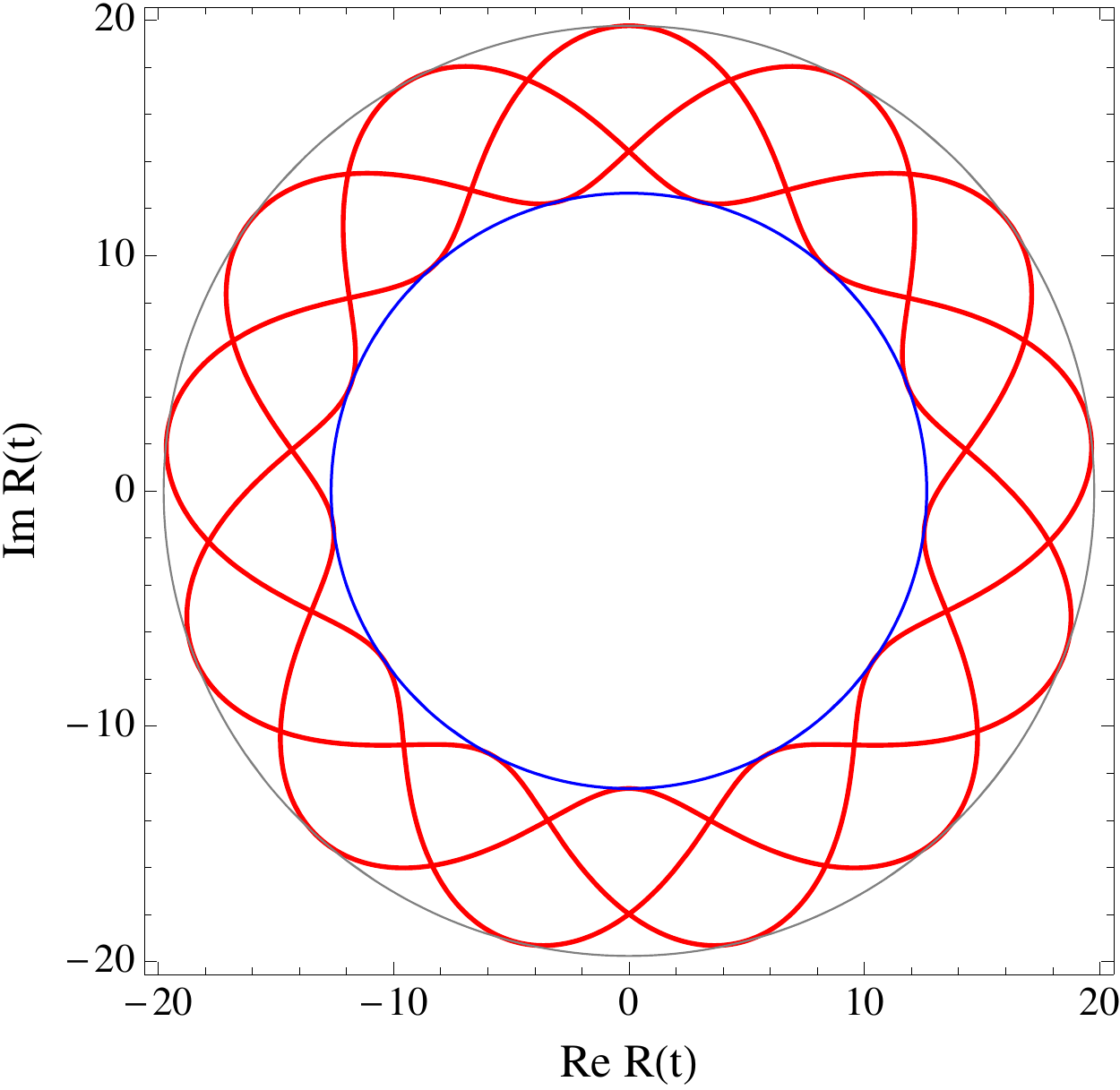}&\hspace*{0.4cm}&\includegraphics[scale=0.45]{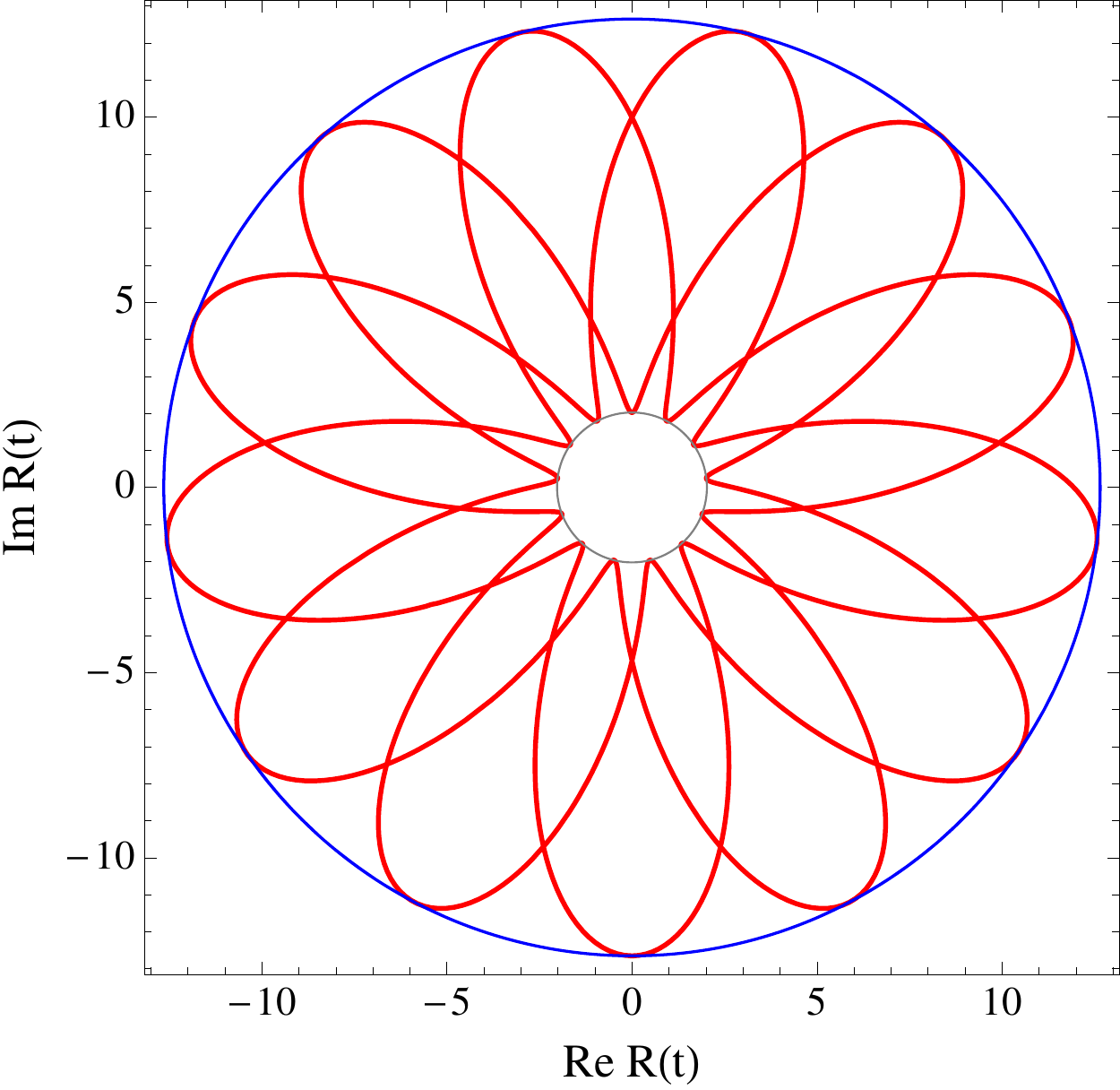}\\
c) & & d)
 
\end{tabular}
\caption{\footnotesize The driving (\ref{rto}) with $g=\sqrt 5$ and $\delta=4$ for (a) $\kappa=0.6$ and (b) $\kappa =3.1$. In the lower graphics $g=\sqrt {160}$ and $\delta=6$, and (c) $\kappa=0.8$ and (d) $\kappa =2.5$. These parameters fulfill the periodicity condition. The blue circle in all the cases corresponds to the case of a circularly polarized field and the gray one is the maximum (minimum) amplitude for $\kappa<1$ ($\kappa>1$).}
\label{RtO}
\end{center}
\end{figure}

\begin{figure}[t]
\begin{center}
\begin{tabular}{ccc}
\includegraphics[scale=0.45]{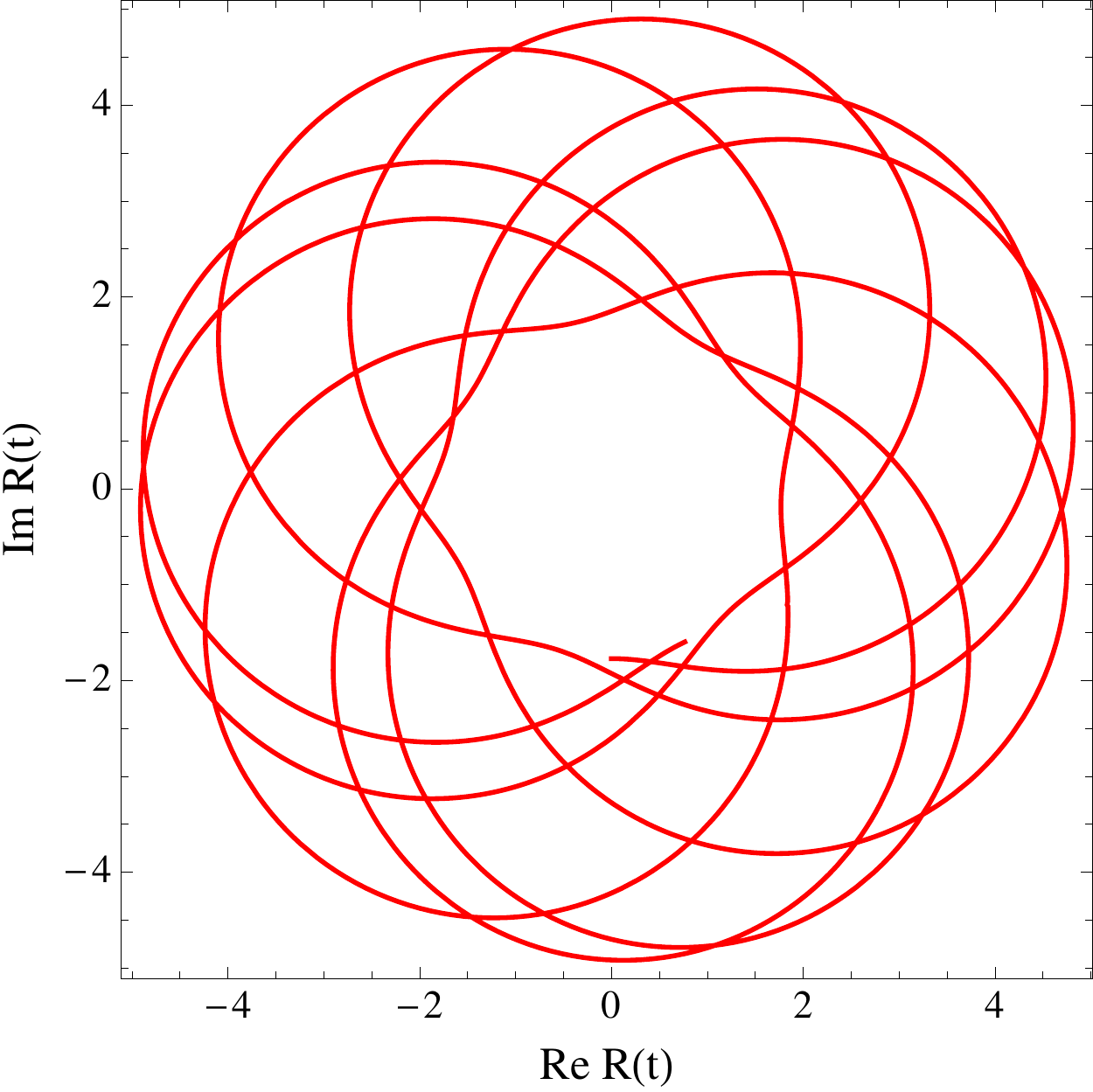}&\hspace*{0.4cm}&\includegraphics[scale=0.45]{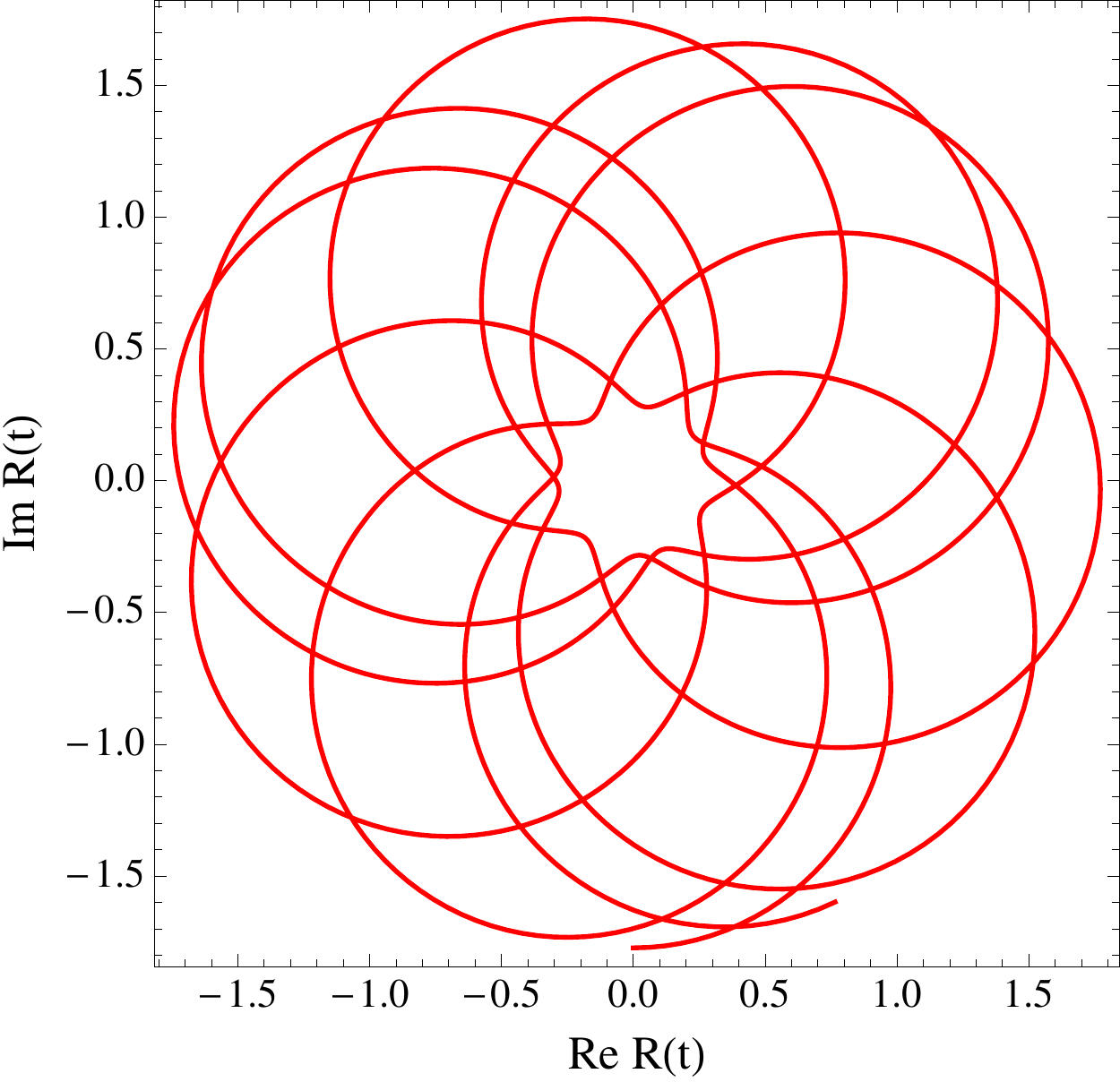}\\
a) & & b)\\
 
\end{tabular}
\caption{\footnotesize The driving (\ref{rto}) with $g=\sqrt \pi$, $\delta=2 \sqrt \pi$ and (a) $\kappa=0.6$ and b) $\kappa=2.5$. These parameters do not fulfill the periodicity condition and the corresponding trajectories are not closed. In both cases the time interval is $[0,10 \kappa \pi/\Omega_0]$}
\label{RtOnp}
\end{center}
\end{figure}

On the other hand, note that the function (\ref{ieta}) is well-defined in the interval $[0,\infty)$ since $\mu^{-2}$ is continous and hence Riemann integrable. 
In Appendix \ref{ap-ieta} we discuss the posibility of writing (\ref{ieta}) in terms of simpler functions and also show that $\eta(\pi/\Omega_1)=\pi/\Omega_0$. This last result is useful to determine the conditions of periodicity of $R$. In fact, the period of the function $\mu$ is $\tau=\pi/\Omega_1$ and according to equation (\ref{pcond}) for $p\in \mathbb{Z}^+$ we must impose the condition that $p\delta/\Omega_0$ is an even integer in order to get a periodic driving field of period $\tau_p=p\tau$, being $p$ the minimum natural number for which the former condition holds.
%
%
The plots of the periodic interactions $R$ are shown in Fig.~\ref{RtO} for several values of $\delta$ and $g$ for which the periodicity condition holds. These describe a flower-like pattern which inscribes (circumscribes) a circle of radius $\vert g\vert$ when $\kappa <1$ ($\kappa >1$). Aditionally, an outer (inner) circle of radius $\vert g\vert /\kappa^2$ bounds the oscillations amplitude. In both cases the number $p$ is related to the number of `petals' of the field, that is to say, the number of times the function $R$ reaches its maximum and minimum values in one period $\tau_p$. On the other hand, in Fig.~\ref{RtOnp} we have chosen the parameters so that the periodicity condition is not fulfilled, giving rise to open curves in the complex plane.
%
%
%

\begin{figure}[t]
\begin{center}
\begin{tabular}{ccc}
\includegraphics[scale=0.5]{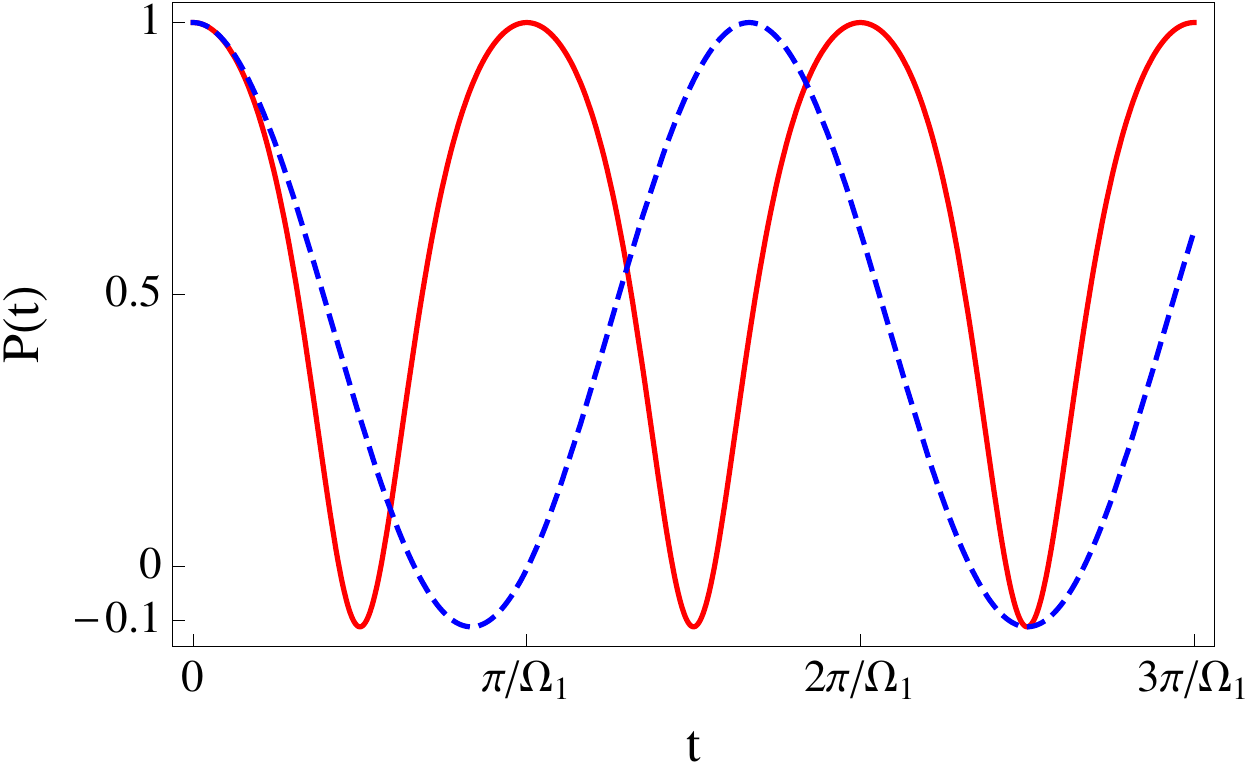}&\hspace*{0.4cm}&\includegraphics[scale=0.5]{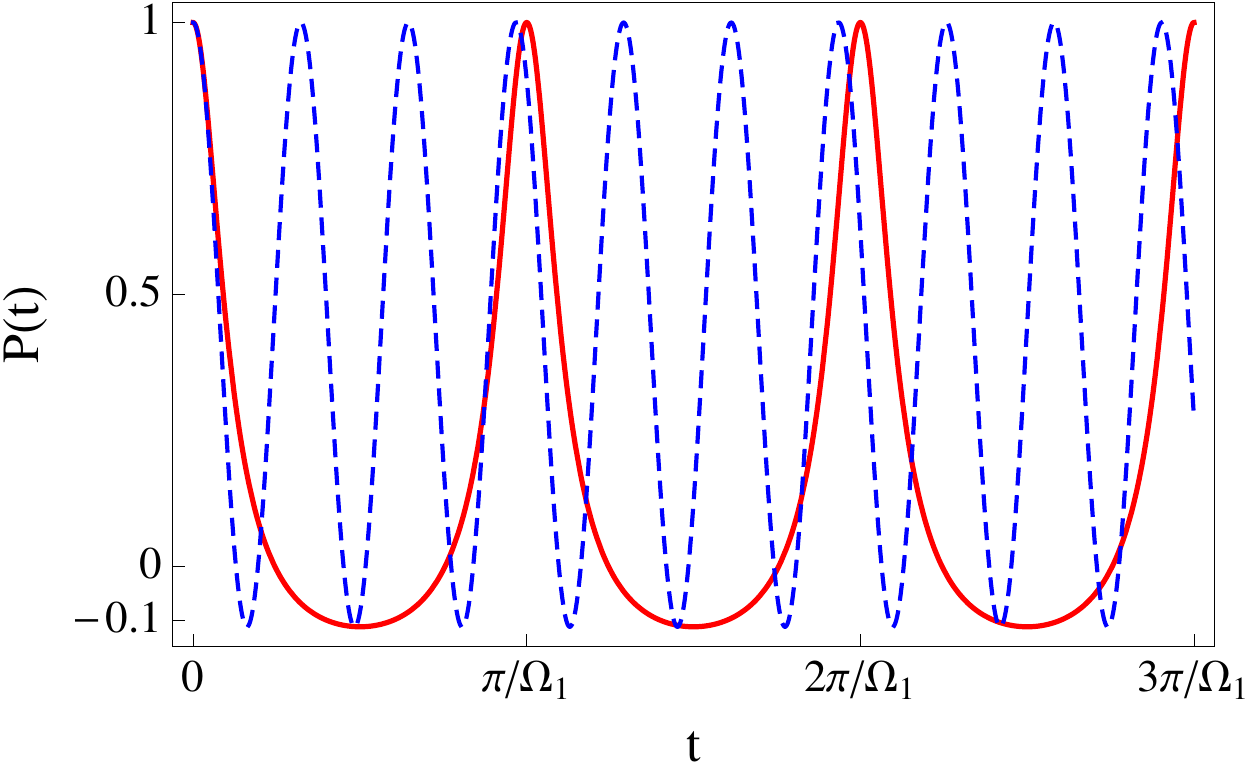}\\
a) & & b)\\[1em] 
\includegraphics[scale=0.5]{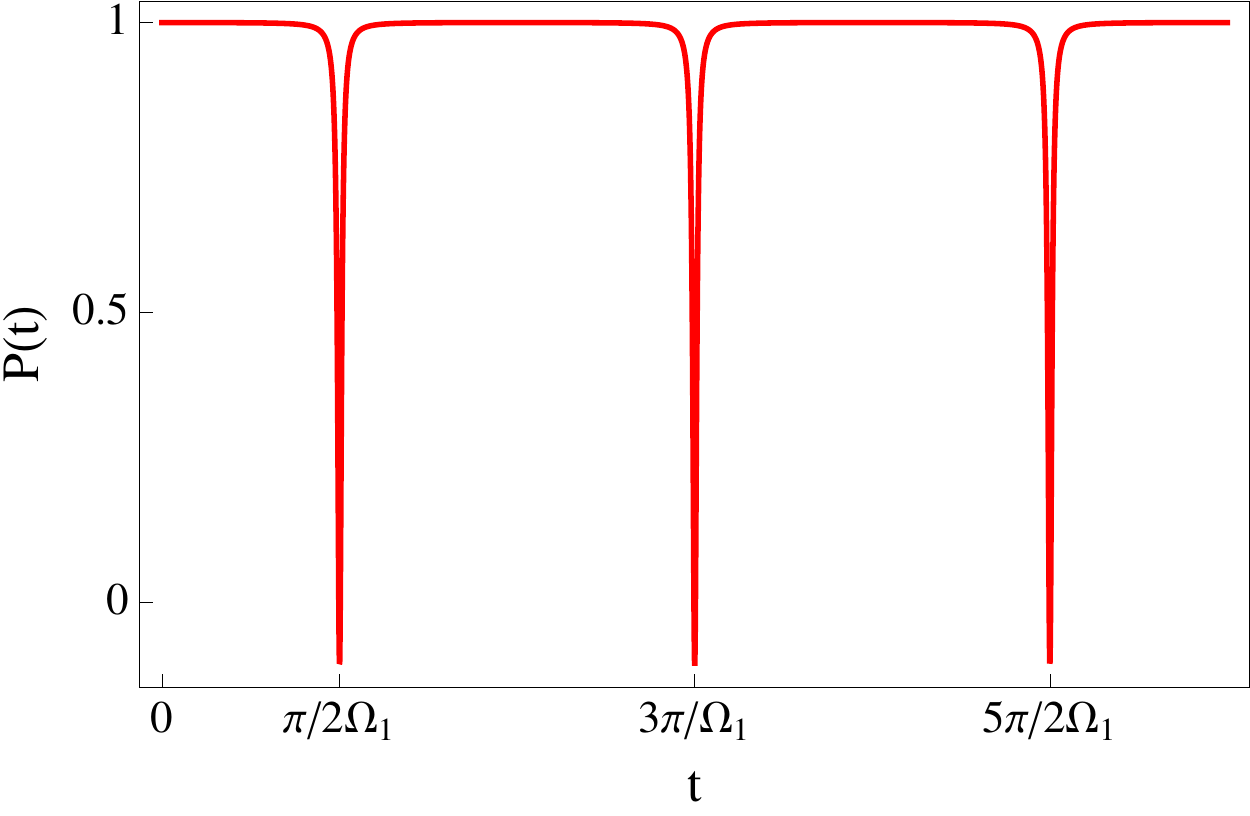}&\hspace*{0.4cm}&\includegraphics[scale=0.5]{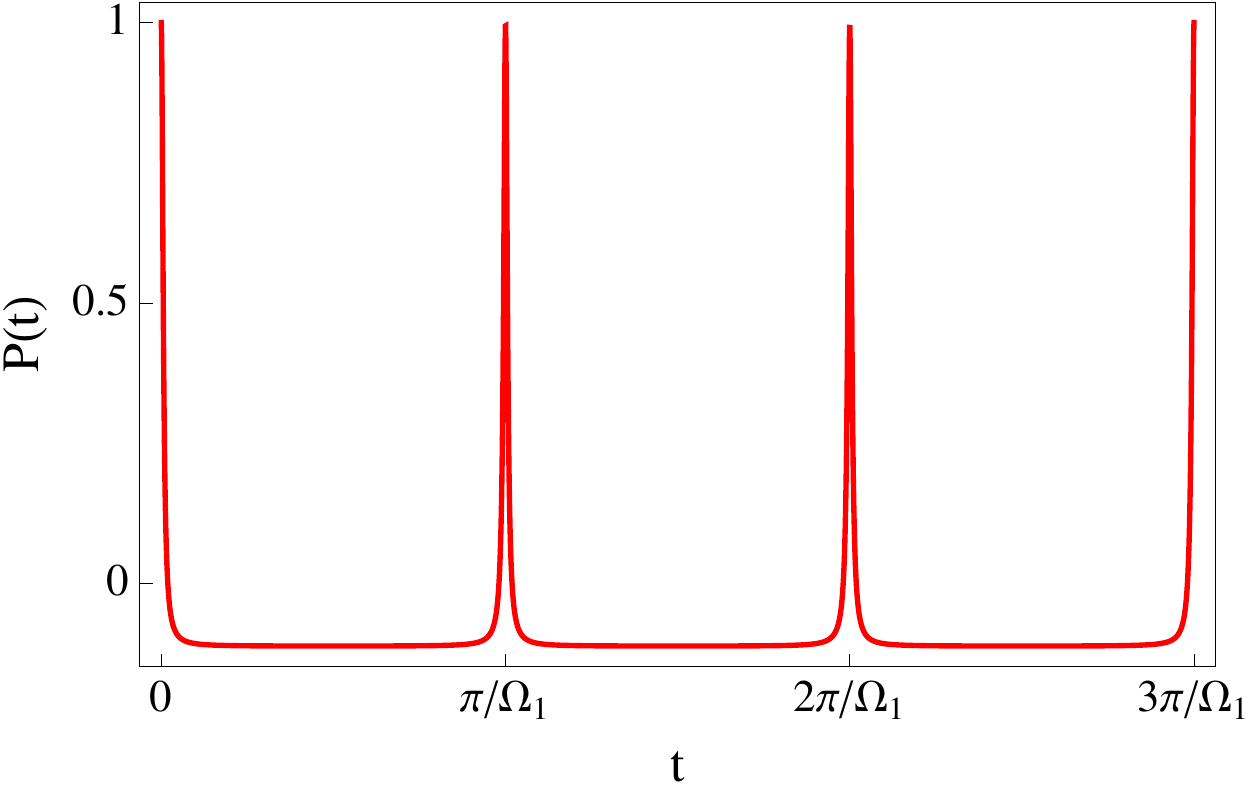}\\
c) & & d)\\ 
\end{tabular}
\caption{\footnotesize Atomic population inversion (\ref{popinv}) as function of time. The dashed line corresponds to the case of a qubit interacting with a circularly polarized field. Besides: a) $\kappa=0.6$ and $\kappa=3.1$ b). The function $P$ shows a colapses and revivals-like behaviour for: c) $\kappa=0.005$ and d) $\kappa=50$
In all cases we have taken $g=\sqrt 5$ and $\delta = 4$.}
\label{RtOpop}
\end{center}
\end{figure}

\begin{figure}[t]
\begin{center}
\begin{tabular}{ccc}
\includegraphics[scale=0.5]{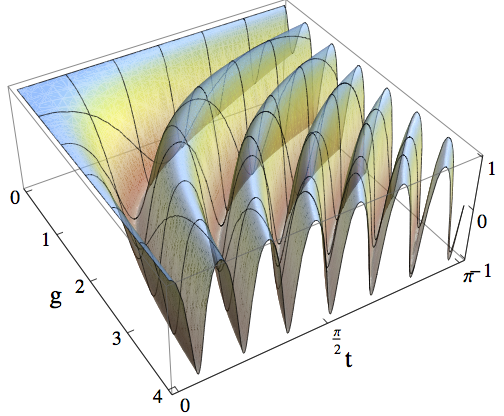}&\hspace*{0.4cm}&\includegraphics[scale=0.5]{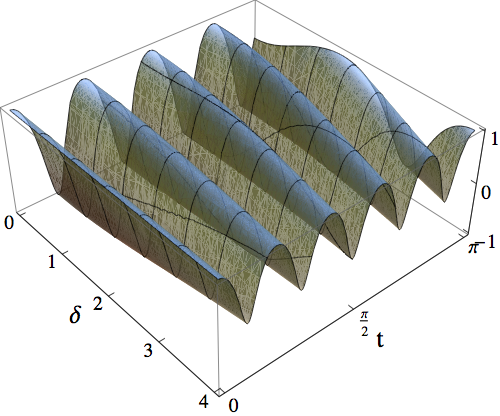}\\
a) & & b)\\ 
\end{tabular}
\caption{\footnotesize Time-evolution of population inversion (\ref{popinv}) as function of the parameters $g$ and $\delta$, which can be used to manipulate the amplitude and period of oscillation.}
\label{s3pop}
\end{center}
\end{figure}

After some calculations we find that the factorizing functions take the form
\begin{equation}
 \alpha(t)=\frac{-2i g \sin(\Omega_1 t) e^{-i\Delta t-i\delta \,\eta(t)}}{2 \Omega_1 \cos(\Omega_1 t)-i\delta \sin(\Omega_1 t)}, \quad  \beta(t)=\frac{-2i {\overline g} \sin(\Omega_1 t)}{2 \Omega_1 \cos(\Omega_1 t)-i\delta \sin(\Omega_1 t)},
\end{equation}
and
\begin{equation}
 \Delta f(t) = \ln\left[\frac{\mu^2(t)}{\varphi^2(t)}\right]-i \delta\,\eta(t)-i \Delta t,
\end{equation}
where the functions $\varphi$, $\mu$ and $\eta$ are given in equations (\ref{phio}), (\ref{muo}), and (\ref{ieta}), respectively. These expressions are useful to describe the time-evolution of the system. For instance, if the initial state is $\vert \psi(0)\rangle=\vert p\rangle$, the state of the system at an arbitrary time is given by
\begin{equation}
\vert\psi(t)\rangle = \displaystyle \frac{e^{-i\Delta t/2-i\eta(t)/2}}{\mu(t)}\left [\cos (\Omega_1 t)+\frac{i\delta}{2 \Omega_1} \sin(\Omega_1t)\right] \vert p\rangle-\frac{i{\overline g}e^{-i\Delta t/2-i\eta(t)/2}}{\mu(t)}\frac{\sin(\Omega_1 t)}{\Omega_1}\vert q\rangle.
\end{equation}
Now it is possible to compute the population inversion as function of time, we get
\begin{equation}\label{popinv}
P(t)=\frac{4 \Omega_1^2\cos^2(\Omega_1 t) +(\delta^2-4\vert g\vert^2)\sin^2(\Omega_1t)}{4\Omega_1^2[\cos^2(\Omega_1 t) +\kappa^2\sin^2(\Omega_1t)]},
\end{equation}
which is a periodic function of time with period $\tau=\pi/\Omega_1=\pi \kappa/\Omega_0$. Remark that for $\kappa=1$ the former expression retrieves the 
conventional expression (\ref{popcp}) for the population inversion of a single qubit driven by a circularly polarized field. One can also note that regardless the value of $\kappa$, the population invertion (\ref{popinv}) has local minima 
\begin{equation}
 P_{\rm min} = \frac{\delta^2-4\vert g\vert^2}{\delta^2+4\vert g\vert^2},
\end{equation}
which are achieved at times $t_n=n \pi/2\Omega_1$, with $n\in \mathbb{Z}^+$. In Fig~\ref{RtOpop} it is presented the population inversion for several values of $g$ and $\delta$ as well as different values of $\kappa$. The blue-dashed curve corresponds to the case of a qubit interacting with a circularly polarized field ($\kappa=1$). It can be also seen that the period of oscillation for $\kappa<1$ is shortened with respect the case $\kappa=1$ while it is enlarged for $\kappa>1$. On the other hand, in the limits $\kappa\approx0$ and $\kappa\gg0$, the oscillations become more defined giving rise to a collapses-and-revivals-like behaviour. This can be noted comparing the shape of the oscillations shown in the plots a) and b) with those depicted in c) 	and d) of Fig.~\ref{RtOpop}, where we have set the value of $\kappa$ according to the limits mentioned before. Finally, the parameters $g$ and $\delta$ can be used to manipulate both the amplitude and shape of the oscillations on demand, as it is shown in Fig.~\ref{s3pop}.



%
\section{Conclusions}\label{conclusions}\label{conc}
We have proposed a procedure to generate exactly solvable families of single qubit driving fields. Requiring that the corresponding time-evolution operator is written as a product of exponential operators we find analytical solutions to the dynamical law using an inverse approach. The control fields are determined by means of the solutions of an Ermakov equation, which are obtained once the time-dependent frequency of the related parametric-oscillator equation is specifyed. The 
dynamics of a qubit interacting with the so-obtained driving fields strongly depends on the form and the parameters of such a function. We have provided some examples of non-periodic and periodic control fields as well as a decaying field.  It has been also shown that the population inversion can be manipulated at will using the parameters the time-dependent frequency depends on. Our results may shed some light in the problem of qubit control and the developement of quantum gates.




\section*{Acknowledgments}

The financial support of CONACyT, Instituto Polit\'ecnico Nacional, M\'exico (Project SIP20170233), the Spanish MINECO (Project MTM2014-57129-C2-1-P) and Junta de Castilla y Le\'on, Spain (VA05U16) is acknowledged. Marco Enr\'iquez is grateful to Departamento de F\'isica, Cinvestav for kind hospitality.


\appendix
\section{The function $\eta(t)$} \label{ap-ieta}
This Appendix is devoted to the search for a primitive of the function $\mu^{-2}$ in $\mathbb{R}^+$. We first consider function defined by the integral in the interval $[0,\pi/2\Omega_1)$
\begin{equation}\label{chi1}
 \chi_1(t)=\int\frac{dt}{\cos^2(\Omega_1t)+\kappa^2 \sin^2(\Omega_1t)}=\frac1{\Omega_0} \arctan[\kappa \tan(\Omega_1t)]+c_1
\end{equation}
the last equality can be easily shown by making the substitution
\begin{equation}\label{utan}
u(t)=\tan(\Omega_1 t).
\end{equation} 
Besides, the initial condition $\displaystyle\lim_{t\rightarrow0^-}\chi_1(t)=\eta(0)=0$ allows us to determine $c_1=0$.
A similar procedure can be accomplished to evaluate an anti-derivative $\chi_2$ for $\mu^{-2}$ in the interval $(\pi/2\Omega_1,\pi/\Omega_1]$. In addition, to ensure the continuity in the point $t=\pi/2\Omega_1$ we must require
\begin{equation}
 \lim_{t\rightarrow(\pi/2\Omega_1)^+} \chi_1(t)=\lim_{t\rightarrow(\pi/2\Omega_1)^-} \chi_2(t),
\end{equation}
from equation (\ref{chi1}) we can evaluate the right-hand of the former condition to obtain $\displaystyle\lim_{t\rightarrow(\pi/2\Omega_1)^+} \chi_1(t)=\pi/2\Omega_0$. We found
\begin{equation}
 \chi_2(t)=\frac1{\Omega_0} \arctan[\kappa \tan(\Omega_1t)]+\frac\pi{\Omega_0}.
\end{equation}
Thus, the the function
\begin{equation}
 \eta(t)=\left\{ \begin{array}{ll} \chi_1(t), & 0\le t<\pi/2\Omega_1 \\[1em]  \pi/2\Omega_0, & t=\pi/2\Omega_1 \\[1em] \chi_2(t), & \pi/2\Omega_1< t\le \pi/\Omega_1 \end{array}\right.
\end{equation}
is a primitive of $\mu^{-2}$ in the interval $(0,\pi/\Omega_1]$. Note that in order to find an antiderivative for this function in $\mathbb{R}^+$ a piecewise function must be defined for each subinterval $[n \pi/\Omega_1,(n+1)\pi/\Omega_1]$ with $n\in\mathbb{Z}^+$ in analogy with the previous analysis.

%

\end{document}